\documentclass[apj,twocolumn]{openjournal}
\pdfoutput=1
\usepackage{graphicx}
\usepackage{enumerate}
\usepackage{amssymb, amsmath, amsfonts}
\usepackage{hyperref}
\usepackage{ulem}
\usepackage{url}
\usepackage{xspace}
\usepackage{xcolor}

\defcitealias{Brandt_2024}{B24}

\usepackage{orcidlink}

\begin{document}

\title{Fitting Moving Objects in Up-The-Ramp Data with Applications to the Roman Space Telescope and JWST}

\author{\vspace{-30pt}Timothy D.~Brandt\,\orcidlink{0000-0003-2630-8073}$^{1}$}
\affiliation{$^1$Space Telescope Science Institute, 3700 San Martin Drive, Baltimore, MD, 21218, USA}
\email{Corresponding author: tbrandt@stsci.edu}

\begin{abstract}

A moving object breaks the fundamental property of constant per-pixel count rates in an astronomical image read out up-the-ramp.  In this paper, we show how to fit a moving object's path across a detector as that detector is read out nondestructively.  We write the full likelihood function for every pixel subject to a constant count rate plus a time-dependent count rate due to a moving source.  Assuming the moving source to be point-like and assuming the effective point-spread function to be known, we are left with four parameters that enter the likelihood nonlinearly: two for position and two for velocity.  All remaining parameters can be optimized using closed-form expressions.  Our approach extracts maximal information on a moving source's position and speed and enables the source to be accurately removed from the image. We investigate the dependence of flux, position, and velocity precision on the target's speed and the readout pattern.  We also find a small, positive bias on the recovered flux due to the need to fit for an uncertain position and speed.  Our approach can be used for space-based images with minor Solar system bodies in the foreground, e.g.~from Roman and JWST, or for ground-based observations with satellites in the foreground.  We demonstrate the promise of our method with a fit to an asteroid track observed serendipitously by the NIRISS instrument on JWST, comparing it to the performance of the JWST pipeline.  Python code implementing our approach is available at \url{https://github.com/t-brandt/moving_source}.  The total computational cost to fit the track of a moving object is $\sim$1 second on a 2023 Macbook Pro.

\end{abstract}



\renewcommand{\topfraction}{.8}
\renewcommand{\floatpagefraction}{.8}

\section{Introduction} \label{sec:intro}

When an astronomical detector is read out many times nondestructively, the count rate for each pixel may be inferred from the individual reads.  The core assumption in this reconstruction is typically that each pixel's count rate is constant: charge accumulates linearly with time \citep{Fixsen+Offenberg+Hanisch+etal_2000}.  In ground-based data, this assumption can break down due to variable atmospheric transmission \citep{Li+McKinnon+Saydjari+etal_2026}, but in space, we expect it to accurately hold for most astronomical scenes. 

The assumption that intensity is constant is no longer true if objects move further than a resolution element on a timescale short compared to an exposure.  This can be the case with minor Solar system bodies in the foreground of a Galactic or extragalactic scene.  An asteroid in the asteroid belt may move at $\sim$10\,km\,s$^{-1}$ relative to Earth at a distance of $\approx$3\,au, for an angular speed of $\sim$10 mas per second, or $\sim$1$''$ over a two-minute-long exposure, much more than the spatial resolution of a telescope like Hubble, JWST, or Roman.  With several successive exposures, an individual asteroid's path may be fit across many images \citep{Burdanov+deWit+Brovz+etal_2025}.  In a detector read out only once, the signal from such a source would be a linear track elongated in the direction of motion.  This track would need to be fit as a starting point for Solar system science \citep{Holler+Milam+Bauer+etal_2018,Holler+Cosentino+Schultz+etal_2025}.

A telescope observing near the ecliptic plane will regularly see minor Solar system bodies in the foreground of astronomical images.  Each pixel will then see a fixed intensity from the background scene plus a time-dependent intensity from each moving foreground object.  For a detector read out nondestructively, the spatially correlated time variation of the count rate can be measured and fit to determine the properties of the moving object and to remove its signal from the image.  On the ground, rapidly increasing numbers of artificial satellites will result in bright sources moving at much higher angular speeds.

This paper addresses the problem of fitting a moving point-like source in a nondestructively-read astronomical image.  We assume that the effective point-spread function, or ePSF \citep{Anderson+King_2000}, is known.  The ePSF describes the image of a point source as sampled by the detector.  We then derive the likelihood of a moving point source atop an unknown, but static, background scene in an image read out up-the-ramp.  This approach can separate the signal of the moving object from the background scene.  It can also constrain the properties---position, velocity, and flux---of the moving object.  We also present the simpler case where the background scene is known and only the moving object itself must be fit.

We organize the paper as follows.  Section \ref{sec:fitramp} derives the pixel-by-pixel likelihood function, shows how to calculate and maximize it, and shows how this may be used to constrain the position, velocity, and flux of a moving object.  Section \ref{sec:synthetic_data} applies the methodology of Section \ref{sec:fitramp} to synthetic data from the Roman Space Telescope's Wide Field Instrument \citep[Roman-WFI,][]{Domber+Gygax+Aumiller+etal_2022,Schlieder+Barclay+Barnes+etal_2024} with read noise, photon noise, and uneven sampling and averaging of reads.  Section \ref{sec:precision_read_pattern} demonstrates the precision achievable with our approach, while Section \ref{sec:biases} discusses biases in the fitted parameters.  Section \ref{sec:implementation} summarizes our Python implementation.  In Section \ref{sec:jwst_data}, we apply our method to fit an asteroid track observed serendipitously by JWST.  We conclude with Section \ref{sec:conclusions}.

\section{Modeling and Fitting a Moving Object} \label{sec:fitramp}

\subsection{The Single-Pixel Likelihood}

We begin by writing the likelihood for an individual pixel subject to both an unknown, but constant, count rate and a time-variable count rate whose time variation is known but whose normalization is not.  In this framework, each pixel has two unknown quantities: a constant count rate, and a coefficient multiplying a normalized, time-variable count rate.  We consider a detector read out nondestructively, but with reads potentially averaged together into groups or resultants.\footnote{JWST uses the term {\it groups}; Roman uses {\it resultants}.  We adopt the term {\it resultants} in this paper.}  This describes the procedure used in all JWST instruments \citep{Jakobsen+Ferruit+Alves+etal_2022,Doyon+Willott+Hutchings+etal_2023,Rieke+Kelly+Misselt+etal_2023,Wright+Rieke+Glasse+etal_2023} and in Roman-WFI \citep{Domber+Gygax+Aumiller+etal_2022,Schlieder+Barclay+Barnes+etal_2024}.

We follow \cite{Brandt_2024}, hereafter \citetalias{Brandt_2024}, and consider the difference between successive resultants scaled by the difference in mean times between resultants.  In \citetalias{Brandt_2024} this value is constant for a given pixel in the absence of noise.  In our scenario it has a constant component and an additive time-variable component for each pixel.  Given a position, velocity, and spatial profile of a moving source, each pixel's count rate due to this source becomes a computable function of time $b \cdot g(t)$ where $b$ is the moving object's flux.  The expected number of counts for each scaled resultant difference $i$ is then $a + b g_i$, where $a$ is the intensity of the background scene and $g_i$ is the expected counts in scaled resultant difference $i$ for a moving source of unit flux.  This framework assumes that the moving source is not itself variable.  For the vast majority of minor Solar system bodies, this is likely to be a good assumption over a timescale of minutes as applicable to a single exposure \citep{Greenstreet+Li+Vavilov+etal_2026}.

In practice the function $g(t)$ is the convolution of an object's intrinsic spatial profile at time $t$ with the effective instrumental point-spread function, or ePSF.  The quantities $g_i$ are defined by
\begin{equation}
    g_i = \langle g(t) \rangle_i
\end{equation}
where the expectation value $\langle \rangle$ is taken between resultants $i$ and $i + 1$, i.e., over the time interval spanned by the mean time of resultant $i$ and the mean time of resultant $i + 1$.   

For the constant (background) scene plus read noise, the covariance matrix of the resultant differences is given by \citetalias{Brandt_2024}.  For the time-variable component, it may be derived using the framework of \citetalias{Brandt_2024} by modifying the adopted times of the individual reads.  Using $b{\cal G}_j$ to denote the total number of counts expected in the pixel through read (not resultant) $j$, we can take the ``time'' of read $j$ to be $b{\cal G}_j$ and apply Equations (8)-(12) of \citetalias{Brandt_2024}.  We then add this covariance matrix to that computed using the actual read times.

Our model is that each pixel's expected number of counts in resultant difference $i$ is $a + b g_i$, where $a$ and $b$ are quantities to be fitted.  Following the notation of \citetalias{Brandt_2024} we define $d_i$ to be the observed counts in the scaled resultant difference $i$ and ${\bf C}$ to be the covariance matrix of the resultant differences.  The $\chi^2$ of our model ($-2$ times the log likelihood) is then
\begin{equation}
\chi^2 = \left({\bf d} - a \cdot {\bf 1} - b \cdot {\bf g} \right)^T {\bf C}^{-1} \left({\bf d} - a \cdot {\bf 1} - b \cdot {\bf g} \right) \label{eq:chisq_first}
\end{equation}
where ${\bf 1}$ refers to a vector of all ones, ${\bf d}$ is the vector of observed resultant differences, and ${\bf g}$ is the vector of $g_i$.  The following subsection is concerned with computing and minimizing Equation \eqref{eq:chisq_first}.

\subsection{Maximizing the Likelihood} \label{subsec:maxlike}

Equation \eqref{eq:chisq_first} describes a quadratic form in $a$ and $b$.  We now turn to the efficient computation of its coefficients.  Following \citetalias{Brandt_2024} we can expand the formula for $\chi^2$ as
\begin{align}
    \chi^2 = &\left( {\bf d}^T {\bf C}^{-1} {\bf d} \right) + a^2 \left( {\bf 1}^T {\bf C}^{-1} {\bf 1} \right) + b^2 \left( {\bf g}^T {\bf C}^{-1} {\bf g} \right) \nonumber \\ &- 2a \left( {\bf 1}^T {\bf C}^{-1} {\bf d} \right) + 2ab \left( {\bf g}^T {\bf C}^{-1} {\bf 1} \right) - 2b \left( {\bf g}^T {\bf C}^{-1} {\bf d} \right) .
    \label{eq:chisq_expanded}
\end{align}
Nearly all of these terms are computed in \citetalias{Brandt_2024}, or may be computed by replacing ${\bf d}$ in \citetalias{Brandt_2024} with ${\bf g}$.  The one exception is the term \begin{align}
        {\bf g}^T {\bf C}^{-1} {\bf d} = \sum_{i=1}^n \sum_{j=1}^n g_i d_j \left({\bf C^{-1}}\right)_{ij} .
\end{align}
For this term, following the notation of \citetalias{Brandt_2024}, we have
\begin{align}
        {\bf g}^T {\bf C}^{-1} {\bf d} = &\sum_{i=1}^n \sum_{j=1}^{i-1} \left(d_i g_j + g_i d_j\right) \left({\bf C^{-1}}\right)_{ij} \nonumber \\
        &+ \sum_{i=1}^n d_i g_i \left({\bf C^{-1}}\right)_{ii} 
\end{align}
which may be written
\begin{align}
        {\bf g}^T {\bf C}^{-1} {\bf d} = &\sum_{i=1}^n \frac{(-1)^i}{\theta_n} \phi_{i+1}\beta_{i-1} \left( g_i (\Theta{\rm D})_{i-1} + d_i (\Theta{\rm G})_{i-1}\right)  \nonumber \\ 
        & + \sum_{i=1}^n d_i g_i \frac{\theta_{i-1}\phi_{i+1}}{\theta_n} 
\end{align}
where all terms are defined in \citetalias{Brandt_2024}, potentially replacing ${\bf d}$ and ${\rm D}$ with ${\bf g}$ and ${\rm G}$.  We can now efficiently compute all of the terms in Equation \eqref{eq:chisq_expanded}, writing that equation as
\begin{equation}
    \chi^2 = {\cal A}_{0} + a^2 {\cal A}_{aa} + 2 ab {\cal A}_{ab} + b^2 {\cal A}_{bb} + 2 a {\cal A}_{a} + 2 b {\cal A}_{b} . \label{eq:chisq_simplified}
\end{equation}

The preceding derivations apply to a single pixel.  When we fit the track of a single moving source, we are fitting the combined response of all pixels.  As such, we must fit all ramps simultaneously.  Holding $b$ fixed, the best-fit static count rate for a given pixel can be found by differentiating and setting the derivative to zero,
\begin{align}
    \frac{\partial \chi^2}{\partial a} = 0 = 2a {\cal A}_{aa} + 2 b {\cal A}_{ab} + 2 {\cal A}_a . 
\end{align}
The best-fit count rate for that pixel is
\begin{equation}
    \tilde{a} = -\frac{{\cal A}_a + b {\cal A}_{ab}}{{\cal A}_{aa}} . \label{eq:a_value_fixed_b}
\end{equation}

To find the best-fit moving source count rate $\tilde{b}$, we sum Equation \eqref{eq:chisq_simplified} over all pixels $k$, substituting Equation \eqref{eq:a_value_fixed_b} pixel-by-pixel.  We have
\begin{align}
    \chi^2_{\rm tot} &= b^2 \sum_k \left( {\cal A}_{bb,k} - \frac{{\cal A}_{ab,k}^2}{{\cal A}_{aa,k}} \right) \nonumber \\
    & ~~~~ + 2b \sum_k \left( {\cal A}_{b,k} - \frac{{\cal A}_{a,k}{\cal A}_{ab,k}}{{\cal A}_{aa,k}} \right) \nonumber \\
    &~~~~+ \sum_k \left( {\cal A}_{0,k} - \frac{{\cal A}_{a,k}^2}{{\cal A}_{aa,k}} \right). \label{eq:chisq_perpixel}
\end{align}
The best-fit $b$, which we denote by $\tilde{b}$, is given by 
\begin{align}
    \tilde{b} &= -\left(\sum_k \left( {\cal A}_{b,k} - \frac{{\cal A}_{a,k}{\cal A}_{ab,k}}{{\cal A}_{aa,k}} \right)\right) \nonumber \\
    &~~~~\times \left(\sum_k \left( {\cal A}_{bb,k} - \frac{{\cal A}_{ab,k}^2}{{\cal A}_{aa,k}} \right) \right)^{-1}
    \label{eq:b_value}
\end{align}
and its standard error is given by
\begin{align}
    \sigma^2_{b} = \left(\sum_k \left( {\cal A}_{bb,k} - \frac{{\cal A}_{ab,k}^2}{{\cal A}_{aa,k}} \right) \right)^{-1} .
\end{align}
We may then substitute Equation \eqref{eq:b_value} into Equation \eqref{eq:a_value_fixed_b} for each pixel to find the best-fit static count rate for that pixel and its final $\chi^2$ value.

We have one final complication: the covariance matrix depends on $b$, but $b$ is not determined until Equation \eqref{eq:b_value}.  This can introduce a bias in the recovered value of $b$ and meaningless $\chi^2$ values. A bias arises because of the covariance between realized and estimated photon noise: pixels with fewer photons simply by chance are estimated to have lower photon noise and thus carry more weight when fitting jointly across pixels (\citetalias{Brandt_2024}).  We avoid this with an iterative approach: we first compute the covariance matrix with $b=0$, compute $b$ via Equation \eqref{eq:b_value}, recompute the covariance matrix with this $b$ value, and finally recompute $b$, $\sigma_b$, and $\chi^2$.

For minor Solar system bodies, a single moving track will almost certainly be sufficient to describe the data.  For artificial satellites it is possible that trails may come in groups that must be fit simultaneously.  Brightnesses might also need to be treated as variable, e.g., as sinusoidal.  In that case Equation \eqref{eq:chisq_simplified} will have cross terms from different objects and/or from modes  of variability of the same object.  The latter could be, e.g., a constant plus a trend plus a sinusoid written as a sine plus a cosine.  The same machinery of this section may be applied to this more complicated case, though the equivalent of Equation \eqref{eq:a_value_fixed_b} will now require the solution of a linear system of equations.  The period of sinusoidal flux variability would enter the likelihood nonlinearly and would have to be treated using the approach of Section \ref{subsec:optimize}.  For the remainder of this paper, we focus on the case of a single moving source of fixed brightness.  

\subsection{The Case of a Known Background}

In some cases, the static astronomical scene may be known independently of a fit to a moving source.  This could be the case if there are repeated observations and a good template exists for difference imaging, or if a field is particularly sparse and the background may be taken to be uniform.  If the static scene is known, its per-pixel count rate may be subtracted from the scaled resultant differences.  The coefficient $a$ in Equation \eqref{eq:chisq_simplified} may then be taken to be zero; ${\cal A}_{ab}$ does not enter into Equation \eqref{eq:chisq_simplified}.  Equation \eqref{eq:chisq_expanded} simplifies to 
\begin{equation}
    \chi^2_{\rm tot} = \sum_k \left( {\cal A}_{0,k} + 2b {\cal A}_{b,k} + b^2 {\cal A}_{bb,k} \right) .
\end{equation}
The best-fit value for $b$, the flux of the moving source, is
\begin{equation}
    \tilde{b} = -\left(\sum_k {\cal A}_{b,k} \right) \left(\sum_k {\cal A}_{bb,k} \right)^{-1} \label{eq:b_knownbg}
\end{equation}
and its uncertainty is given by
\begin{equation}
    \sigma^2_b = \left(\sum_k {\cal A}_{bb,k} \right)^{-1}. \label{eq:err_b_knownbg}
\end{equation}
If a uniform background is to be fit, the $\chi^2$ values for each pixel, Equation \eqref{eq:chisq_simplified}, may be added together retaining the $a$ terms.  The result will be a quadratic form in two parameters---the background level and the flux of the moving source---that can be optimized simultaneously.  Without the need to disentangle the contributions from an unknown static scene and a moving source, the precision on the moving source can be much better than in the more general case.

\subsection{A Model of the Illumination} \label{subsec:illum_model}

We now turn to the problem of deriving the illumination function $g(t)$ of each pixel from an assumed sky path.  We begin by assuming that we have the oversampled image of the moving object as it would appear if it were stationary.  This could be either an ePSF or an ePSF convolved with an intrinsic spatial profile (e.g.~a circle or a Lambertian).  The integrated light seen by a pixel between two reads is an integral of this oversampled source image along the direction of motion.  For this section, and for the rest of the paper, we assume that the cadence of the detector readout is constant.  This is indeed the case for all detectors on JWST and for the eighteen detectors of Roman-WFI.

The problem of integrating the oversampled image along a given direction is easier if that direction is parallel to one of the axes of the grid on which this image is defined.  In practice, we may rotate the oversampled image so that one of the axes is aligned with the direction of motion, perform this integral as a convolution to compute the image as smeared over a single read, and then rotate back.  If we use a bicubic spline to represent the image in the frame parallel to a detector axis, the integral may be performed exactly.  

Along one of the grid axes, a bicubic spline is simply a cubic spline, a piecewise cubic polynomial of the form 
\begin{equation}
    z_k(x_k) = \sum_{i=0}^3 a_{i, k} x_k^i
\end{equation}
where $x_k$ is the distance from the nearest grid point $k$ and $a_{i,k}$ are coefficients defined on that interval.  If we wish to compute the image as it would appear smeared out by constant motion over a distance $vt$ in the $x$-direction, for a pixel initially sampling $x_k=0$, we may evaluate
\begin{align}
    z'_k(x_k, vt) &= \frac{1}{vt} \int_0^{vt} z_k(x_k) d x_k \nonumber \\
    &= \sum_{i=0}^3 \frac{1}{i + 1} a_{i,k} \left(vt \right)^i .
    \label{eq:smearing_equation_first}
\end{align}
The coefficients $a_{i,k}$ themselves may be straightforwardly computed within each grid interval from the cubic spline representation.

Equation \eqref{eq:smearing_equation_first} is valid for motion of at most one pixel per read, i.e., $vt \leq 1$.  If $vt > 1$, we may treat the integer part of $vt$ and the fractional part separately.  We obtain
\begin{align}
    z'_k(x_k, vt) &= \sum_{j=k}^{m - 1} \sum_{i=0}^3 \frac{a_{i,j}}{i + 1} + \sum_{i=0}^3 \frac{a_{i,m}}{i + 1} (vt - \lfloor vt \rfloor)^{i}
    \label{eq:smearing_equation}
\end{align}
with $m = k + \lfloor vt \rfloor$ and $\lfloor\rfloor$ denoting the floor operator.  The $z'_k$ values along this grid axis, together with the corresponding values from all other grid axes, form the desired image smeared out by constant motion, and may be interpolated to obtain the total number of counts received by a given pixel between two reads.  

We demonstrate this procedure using a synthetic ePSF of the Roman Space Telescope's Wide Field Instrument (Roman-WFI) as available on the Calibration Reference Data System (CRDS) at the time of writing\footnote{\url{https://roman-crds.stsci.edu}}; we use the ePSF at the center of Detector 1.  As of the time of publication, the PSF on CRDS has not been convolved with the pixel response function so we perform a convolution with a top-hat pixel response.  Our reference ePSF oversamples the detector pixels by a factor of four.  This ePSF should accurately describe the image of a source small compared to Roman-WFI's resolution of $\approx$100\,mas ($\approx$70\,km at 1\,au).

\begin{figure*}
\begin{center}
    \includegraphics[width=0.6\textwidth]{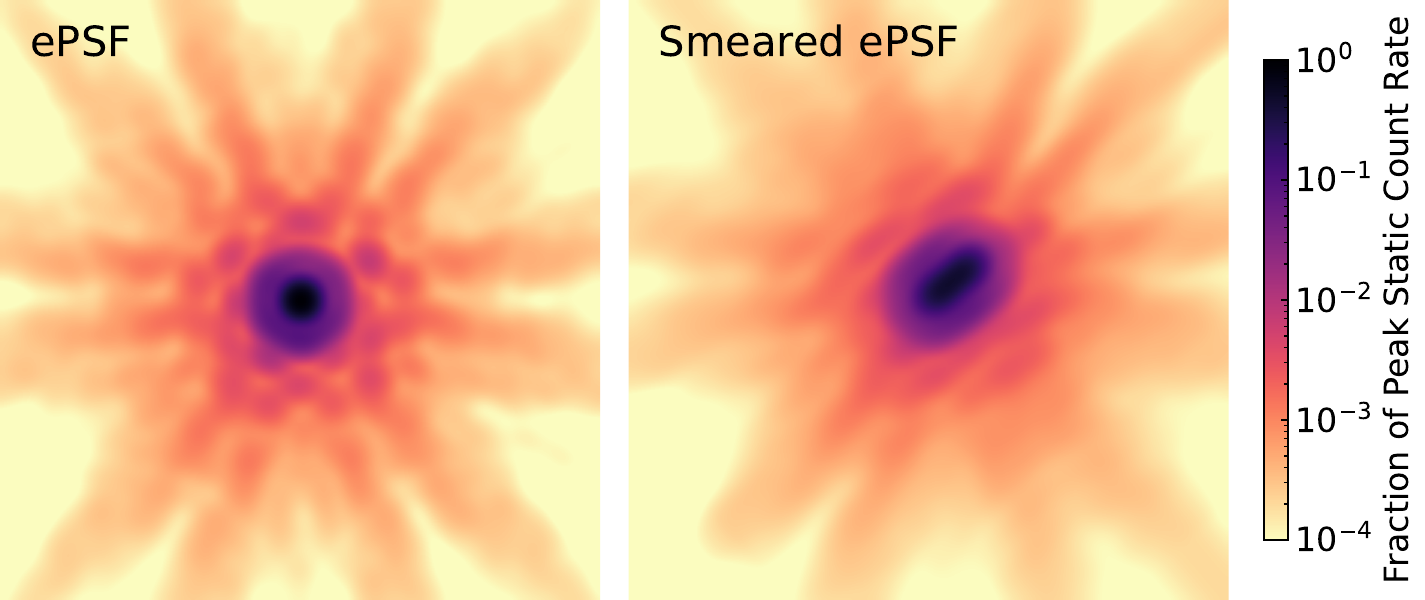} \\[2em]   
    \includegraphics[width=\textwidth]{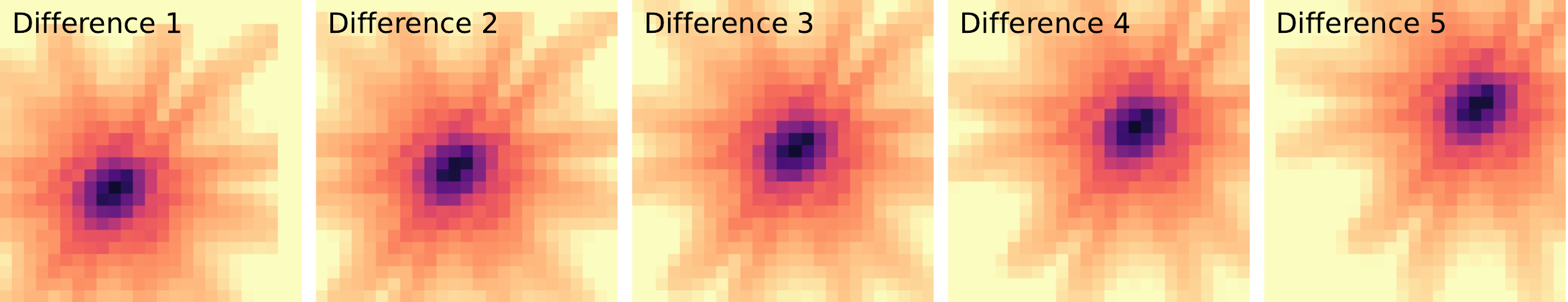} 
\end{center}
\caption{Top-left: the assumed ePSF.  Top-right: the same ePSF integrated along a line as described in Section \ref{subsec:illum_model}.  Bottom sequence: smeared ePSFs as in top-right sampled at the pixels corresponding to successive read differences.  The direction and extent of motion match the direction and extent of smearing.  Edge effects are visible from the finite size of the ePSF used for this computation. \label{fig:ePSF_smearing}}
\end{figure*}

The top panel of Figure \ref{fig:ePSF_smearing} shows the ePSF on the left as taken from CRDS and convolved with a top-hat pixel response function.  The top-right panel shows this ePSF smeared over 2.7 pixels at an angle 50$^\circ$ clockwise from vertical.  As noted in this section, we perform this smearing by rotating the ePSF so that its grid axes align with the direction of motion.  We then use the spline representation to perform the integral exactly and rotate back.  We adopt a logarithmic color scale normalized to the peak value of the ePSF.  We limit the computation to an ePSF of size $32 \times 32$ detector pixels.  This represents a compromise between computational performance and residual edge effects from a finite ePSF size. 

The bottom panel of Figure \ref{fig:ePSF_smearing} shows a sequence of templates in which the smeared ePSF of the top-right panel is interpolated at different positions.  The source in this case is moving at 2.7 pixels/read, so in each subsequent read, its center is displaced by 2.7 pixels along the direction of motion.  The template in each read difference consists of a single interpolation of the smeared ePSF.  The color scale for this sequence is the same as that for the top panel.

Edge effects from the finite size of the adopted ePSF may be seen in the first and last read differences as the ePSF cuts off to zero.  These edge effects could be reduced by adopting a larger ePSF at the cost of additional computational effort.  While the computational cost is negligible when making a demonstration image like that shown in Figure \ref{fig:ePSF_smearing}, it becomes significant when fitting many tracks of moving objects.

\subsection{Optimizing over Position, Velocity, and Shape} \label{subsec:optimize}

The bottom panel of Figure \ref{fig:ePSF_smearing} shows a sequence of template images in individual read differences.  We may perform a cumulative sum of these templates to obtain the integrated counts seen by each pixel as a function of read number.  Detectors that read out up-the-ramp may or may not save each individual read.  They may also average reads together into groups (the JWST terminology) or resultants (the corresponding term for Roman), and they may drop some reads entirely.  Starting with the integrated counts observed in each read, we may process our templates in the same way as the actual reads.  These processed templates may then be used as ${\bf g}$ in Equation \eqref{eq:chisq_expanded}.  

Assuming a moving object to be spatially unresolved, our template has four free parameters: two for position, and two for velocity.  The best $\chi^2$ value will depend on ${\bf g}$, and hence, on these four parameters.  This constitutes a nonlinear optimization problem.  We use {\tt scipy.optimize.minimize} with the Nelder-Mead algorithm \citep{Nelder+Mead_1965} to find the best position and velocity.  If we wish to further constrain the moving object's shape, we may apply this shape as a convolution to the ePSF shown in Figure \ref{fig:ePSF_smearing}.  There will then be one or more additional parameters to optimize. 

\section{Demonstration on Synthetic Data} \label{sec:synthetic_data}

We now create synthetic data to demonstrate the approach of the previous section.  We adopt the same Roman-WFI model ePSF as that shown in Figure \ref{fig:ePSF_smearing}; it oversamples the detector pixels by a factor of four.  We take the same angle of motion as in Figure \ref{fig:ePSF_smearing}, 50$^\circ$ clockwise of vertical, but use a speed of 1.6 pixels/read.  For this investigation we use the {\tt IM\_107\_7} multiaccumulation table intended for Roman's High Latitude Wide Area Survey (HLWAS)\footnote{\url{https://roman-docs.stsci.edu/roman-instruments/the-wide-field-instrument/observing-with-the-wfi/wfi-multiaccum-ma-tables/imaging-multiaccum-tables\#IM_107_7}}.  This readout pattern consists of 32 reads averaged together into six resultants.  The six resultants consist of 1, 2, 8, 16, 4, and 1 read(s), respectively.  

\begin{figure*}[!ht]
    \includegraphics[width=\textwidth]{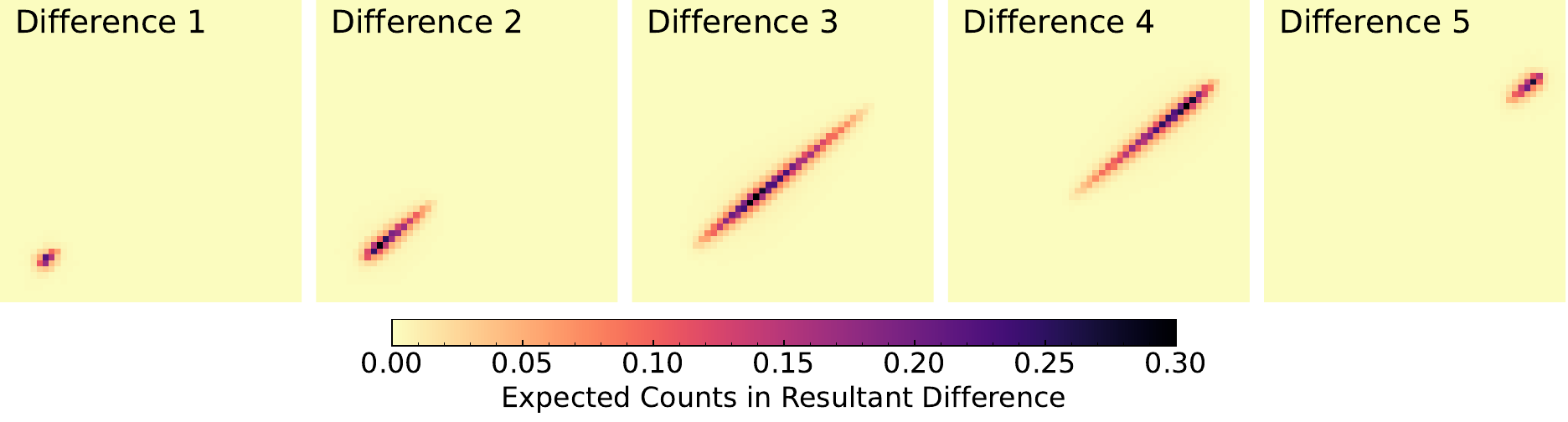}
    \caption{Expected counts in every pixel of each resultant difference for a point source of unit flux (1 count/read in total) moving at a speed of 1.6 pixels/read.  The six resultants consist of 1, 2, 8, 16, 4, and 1 read, respectively, for 32 reads in total. \label{fig:modelresultants_hlwas}}
\end{figure*}

Figure \ref{fig:modelresultants_hlwas} shows the expected counts in each pixel in each resultant difference for a moving source with an integrated count rate of 1 count/read.  The first resultant difference consists of the difference between four reads and one read and shows relatively modest smearing.  The third and fourth resultant differences both include the long fourth resultant, which consists of 16 reads (half the total exposure time).  The resulting streaks are far longer than those seen in the last and especially the first resultant difference.

Next, we generate synthetic data using this template for a moving source.  We consider two scenarios: one in which the background is uniform, and one in which there is a small uniform background together with a number of static point sources.  In the former case we adopt a count rate for the background of 5 counts/read and a count rate for the moving source of 500 counts/read.  For the latter case we use a uniform background of 1 count/read and add 100 stars with random positions and random fluxes ranging from 0 to 30 counts/read.  

We generate photon noise and read noise realizations read-by-read, assuming the units of counts to be electrons and adopting a read noise of 10\,e$^-$/read.  We then group these noisy measurements into resultants and attempt to recover the parameters of the moving source.  We begin with a good, but not an exact, initial guess for its position and velocity.  Our use of the Nelder-Mead algorithm requires a good initial guess to reliably converge.  In practice this would need to be provided by a separate, more approximate analysis to identify the likely locations and approximate parameters of moving objects.  We defer the problem of identifying regions of interest and obtaining good initial guesses to future work.  Once we have a good initial guess, we minimize the $\chi^2$ of Equation \eqref{eq:chisq_expanded} by optimizing the position and velocity of the moving object.  Its flux is optimized as part of the $\chi^2$ minimization discussed in Section \ref{sec:fitramp}.  

\begin{figure*}
    \begin{center}
    \includegraphics[width=0.8\textwidth]{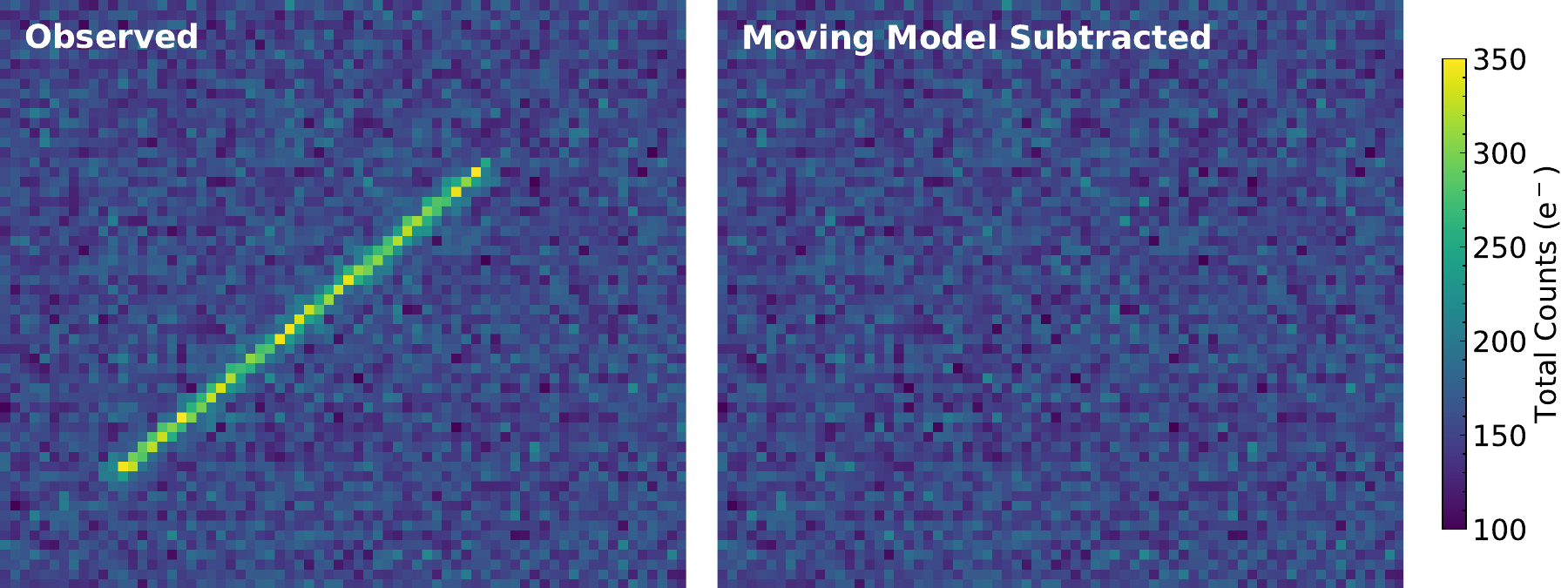} \\[2em]
    \includegraphics[width=0.8\textwidth]{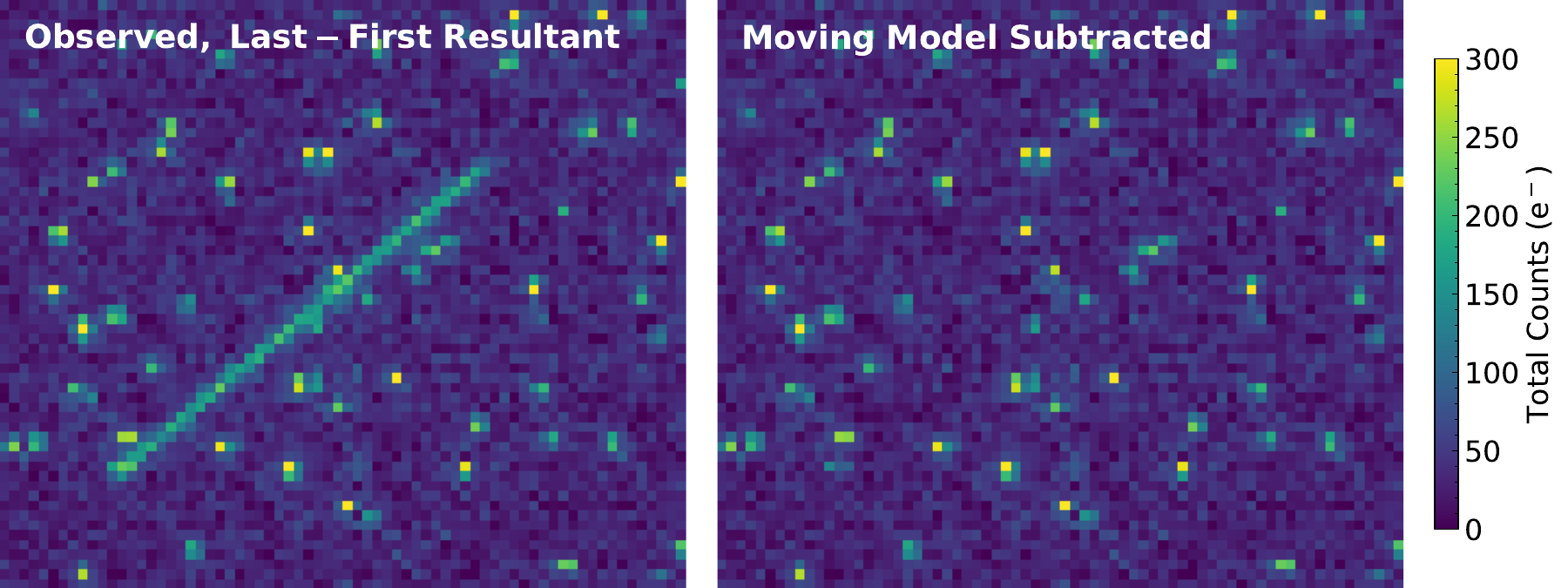}
    \end{center}
    \caption{Left panels: the total number of electrons for a static scene plus a moving point source over 32 reads including both photon noise and read noise (10\,e$^-$/read).  In the top panel, the static scene is a constant 5\,e$^-$/read, while in the bottom panel, the scene is a spatially uniform 1\,e$^-$/read plus 100 point sources with random positions and brightnesses.  Right panels: the same data as in the left panels after jointly fitting pixel-by-pixel intensities and the parameters and flux of a moving object, and removing the best-fit moving point source model.  \label{fig:exampleskypaths}}
\end{figure*}

Figure \ref{fig:exampleskypaths} shows our results.  The left panels show the integrated counts over all 32 reads, including the effects of read noise and photon noise.  The uniform background case is shown on top while the case of the background star field is shown on the bottom.  The streak from the moving object is the sum of all panels of Figure \ref{fig:modelresultants_hlwas}, scaled to our adopted flux and with the addition of photon noise.  The right panels of Figure \ref{fig:exampleskypaths} are the same as the left panels, but with a model sky path computed from the best-fit position and velocity of the moving source and subtracted off.  No assumptions have been made about the spatial structure of the static scenes.  The streaks from the moving objects are no longer visible, though we defer a quantitative discussion of biases in the fitting to Section \ref{sec:biases}.  In the remainder of this section we compare the approach of this paper to standard pipeline reductions, and we show the behavior of the per-pixel value of the $\chi^2$ goodness-of-fit metric. 

\subsection{Comparison to Standard Pipeline Approaches}

A standard data reduction pipeline, like the JWST pipeline \citep{JWST_pipeline_2023} or {\tt romancal}\footnote{\url{https://github.com/spacetelescope/romancal}}, will not properly process a moving object.  The track of a moving object as processed through these pipelines will differ substantially from that seen in either panel of Figure \ref{fig:exampleskypaths}, complicating any identification and fitting in images processed into best-fit count rates.  This is for two main reasons: the weightings of the individual resultants in the fit, and the impact of jump detection designed to flag and remove cosmic rays.

\begin{figure*}[!ht]
    \includegraphics[width=\textwidth]{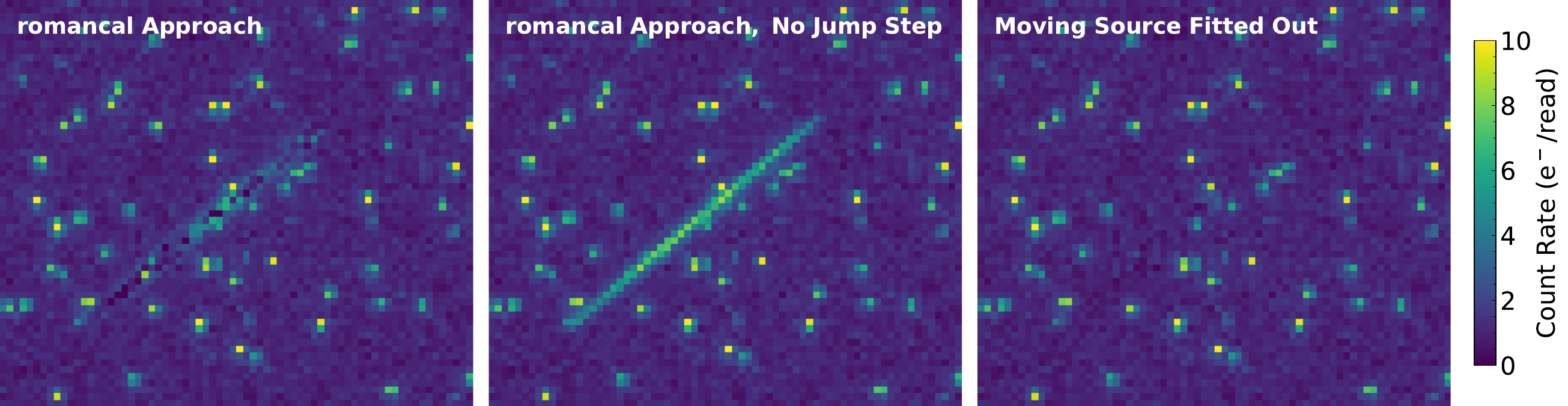} \\[1em]
    \includegraphics[width=\textwidth]{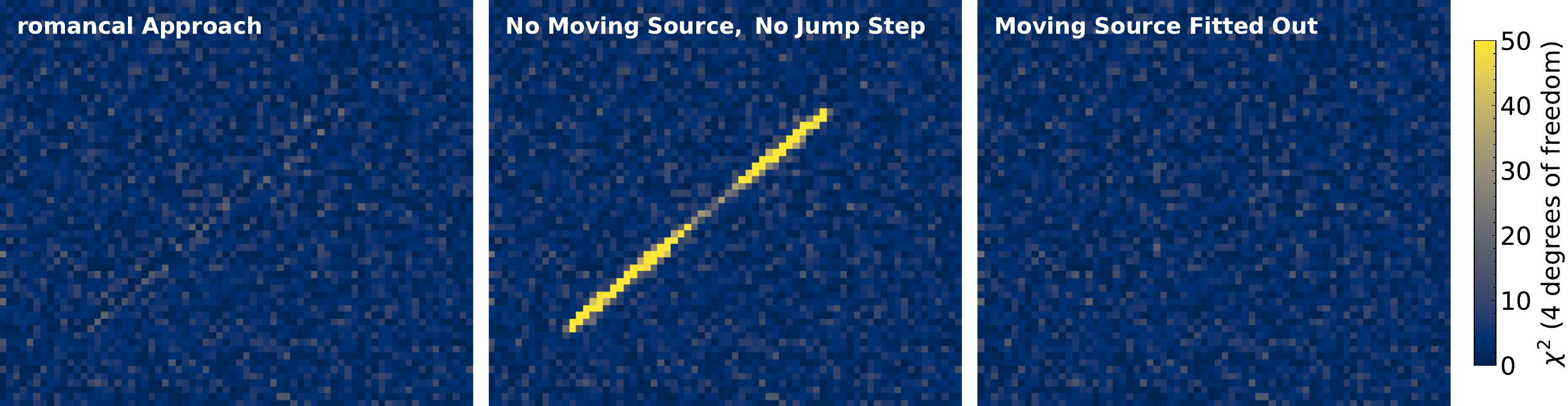}
    \caption{Top row: comparison between best-fit rate images obtained using the jump detection of \cite{Brandt_2024b} and ramp fitting of \citetalias{Brandt_2024} (left); ramp fitting of \citetalias{Brandt_2024} without jump detection (center); and the static count rate using the approach of this paper to fit the moving object's track (right). Bottom row: $\chi^2$ values (Section \ref{subsec:chisq}) corresponding to the three top panels.  Many pixels in the bottom-middle panel have $\chi^2$ values far exceeding the color bar's upper limit of 50. \label{fig:romancal_comparison}}
\end{figure*}

The top row of Figure \ref{fig:romancal_comparison} shows the same moving object track as that shown in the bottom panels of Figure \ref{fig:exampleskypaths}, but processed into a count rate three ways.  The top-left panel uses the jump detection algorithm in \cite{Brandt_2024b} combined with the ramp fitting approach in \citetalias{Brandt_2024}, while the top-middle panel omits the jump detection step and uses all resultants in the ramp fit.  The image shown at top-right incorporates the full methodology of this paper, showing the best-fit static count rate for each pixel.  The top-left panel shows much reduced count rates in the center of the track, while the top-middle panel shows the ends of the track faded relative to Figure \ref{fig:exampleskypaths}.

The differences between the top-left and top-center panels of Figure \ref{fig:romancal_comparison} are due to the impact of jump detection.  Jump detection is intended to mitigate the effects of cosmic rays, which can deposit large amounts of charge on a pixel almost instantaneously.  Pipelines for space-based observatories like JWST and Roman attempt to flag group or resultant differences that are affected by jumps and remove them from the fit \citep{Sharma+Casertano_2024,Brandt_2024b}.  For a single pixel, a moving source will induce behavior similar to that of a cosmic ray: a sharp jump in counts as the track of the moving object passes over a pixel.  Jump detection algorithms will therefore remove much of the light from a moving object if it is sufficiently bright, especially along the core of the track.  The very center of the track in this example corresponds to a long resultant with a relatively large contribution of read and photon noise from the static scene.  Jump detection is less sensitive as a result: several pixels in the middle of this track have not had any resultant differences masked by jump detection.

The difference between the top-middle panel of Figure \ref{fig:romancal_comparison} and the left panel of Figure \ref{fig:exampleskypaths} is due to the weighting of resultants in a slope fit.  Where photon noise dominates, only the first and last reads contribute to the fit, and the best-fit slope matches that shown in Figure \ref{fig:exampleskypaths}.  Where read noise dominates, the middle resultants contribute increasingly to the fit.  When counts do not accumulate at a constant rate, this leads to differing inferred count rates depending on when in a ramp those counts are recorded.

The combination of effects shown in the two left panels of the top row of Figure \ref{fig:romancal_comparison}, from jump detection and the weighting of resultants, precludes any straightforward fitting of moving objects in the best-fit rates even if the static scene is known.  Any approach to modeling a moving source must operate on the values of the resultants themselves.

\subsection{$\chi^2$ and the Goodness of Fit} \label{subsec:chisq}

The likelihood as written in Section \ref{subsec:maxlike} gives the best-fit $\chi^2$ value for every pixel via Equation \eqref{eq:chisq_perpixel}, substituting in the value of $b$ given by Equation \eqref{eq:b_value}.  If the model for the count rate in that pixel, including the covariance matrix, is accurate, this value should be nearly $\chi^2$-distributed with degrees of freedom equal to the number of resultant differences minus one (for the count rate that must be fitted); see \citetalias{Brandt_2024} for a discussion.  In our case, with six resultants, we have four degrees of freedom per pixel.  There are five additional parameters for the moving object itself.  However, assuming the number of pixels affected by the moving object's track to be $\gg$5, the impact of these extra parameters on any given pixel is small.

The bottom row of Figure \ref{fig:romancal_comparison} shows the $\chi^2$ values for each pixel in the corresponding panel above.  For the panel at the lower-left, we fit a constant per-pixel count rate using the approach in \citetalias{Brandt_2024}, but jumps are inferred as in \cite{Brandt_2024b} and omitted from the fit.  This results in reasonable $\chi^2$ values overall, and especially along the track center where the contribution of the moving source is flagged as a series of jumps and ignored in the fit.  The $\chi^2$ map has elevated values---poorer fits---adjacent to the center of the track where the moving object is not quite bright enough to trigger jump detection.  The lower-middle panel omits jump detection entirely and fits a constant per-pixel count rate.  The $\chi^2$ values in this case are very poor, as a constant count rate does not describe the signal seen by a pixel in the center of the track.  The $\chi^2$ values in this case can exceed 100 for four degrees of freedom, a statistical near-impossibility if the underlying model were correct.  The lower-right panel of Figure \ref{fig:romancal_comparison} shows the $\chi^2$ values when fitting the correct statistical model to the data, i.e., using the approach described in this paper.  

Figure \ref{fig:romancal_comparison} demonstrates that the per-pixel $\chi^2$ values retain their usefulness even in the presence of a moving object.  We can use them to flag candidate moving objects in data by the goodness of fit of constant per-pixel count rates.  We can then further use the per-pixel $\chi^2$ to evaluate the goodness-of-fit of the track of a moving object with the expectation of a formally good fit.  

\subsection{Oblique Cosmic Rays}

\begin{figure*}[!ht]
    \includegraphics[width=\textwidth]{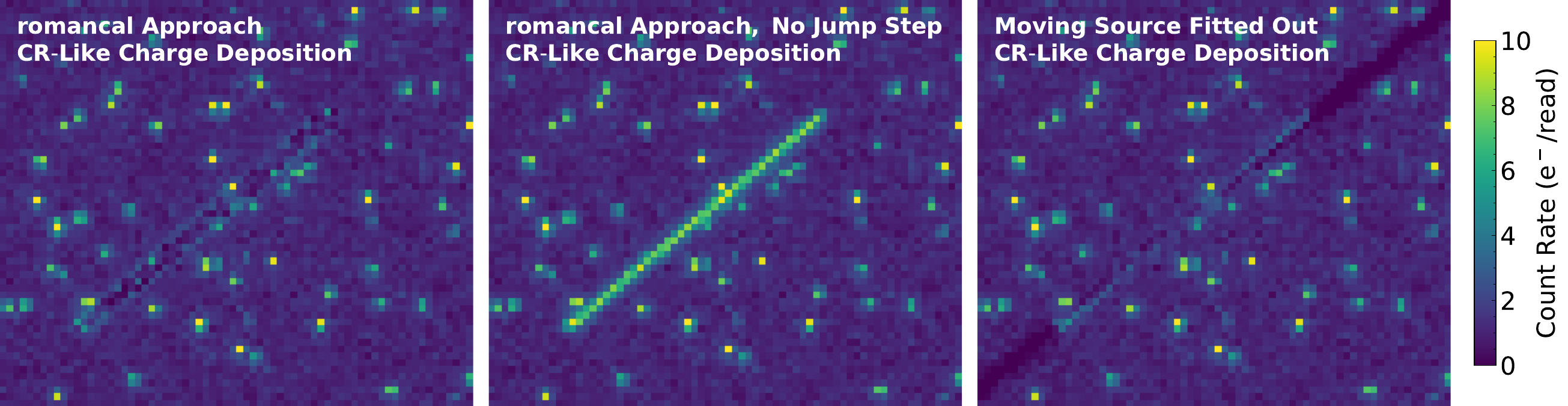} \\[1em]
    \includegraphics[width=\textwidth]{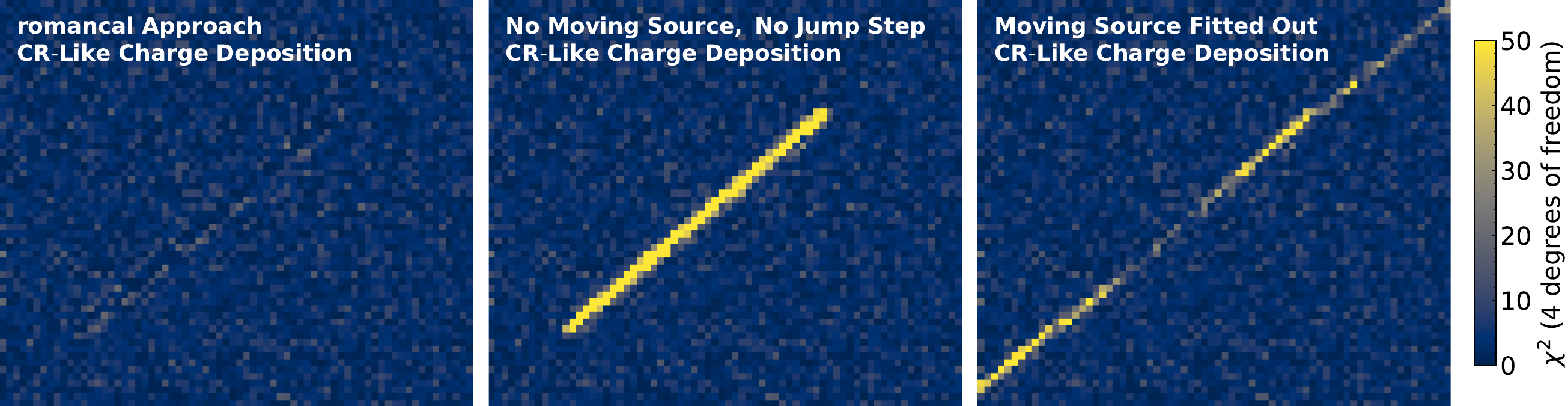} 
    \caption{Results when fitting the same track as that shown in Figure \ref{fig:exampleskypaths}, but with the charge deposited all at once rather than spread out over time.  All panels use the same processing as in Figure \ref{fig:romancal_comparison}; the bottom panels show the per-pixel $\chi^2$ values corresponding to the count rates and fits in the upper panels.  With all excess charge deposited at once, a moving source model does not describe the data and provides a poor fit.  The top-right panel shows a negative track extending past the track of the true moving object, while the lower-right panel shows many pixels with very poor $\chi^2$ values throughout the track and beyond (c.f.~lower-right panel of Figure \ref{fig:romancal_comparison}).  \label{fig:oblique_CR}}
\end{figure*}

A cosmic ray hitting the detector will deposit charge; a cosmic ray at an oblique angle can deposit charge in a line.  This looks superficially similar to the track of a moving object, especially if we consider only the difference between the last and the first resultant.  A significant difference is in the temporal character of the charge deposition: it happens at once, rather than incrementally over the exposure.  

In this section we take the same track and background scene as that shown in the lower panel of Figure \ref{fig:exampleskypaths}.  However, instead of depositing the charge over the entire ramp, we add it all between Resultants 3 and 4.  In analogy to Figure \ref{fig:modelresultants_hlwas}, we take the sum of the five panels of that figure and place it all in Difference 3; we set the other resultant differences to zero.  The difference between the first and last resultants, the lower-left panel of Figure \ref{fig:exampleskypaths}, is unchanged from the case of a true moving object.  The outcome of fitting a moving object's time-dependent sky path, however, is very different.

Figure \ref{fig:oblique_CR} shows the results of fitting a moving object's track, both in the inferred static count rate (top) and in the $\chi^2$ of the fit (bottom).  The top-left panel uses jump detection in addition to the ramp fitting approach of \citetalias{Brandt_2024}.  The core of the track is masked more effectively than in the left panel of Figure \ref{fig:romancal_comparison}, as the excess charge deposition now follows the assumption of the jump detection algorithm.  The fainter parts of the track, however, persist.  The track also looks similar to the corresponding (center) panel of Figure~\ref{fig:romancal_comparison} if processed without jump detection.  In contrast to the right panel of Figure~\ref{fig:romancal_comparison}, a moving object model (right) provides a poor fit.  The moving object fit removes much of the signal from the part of the ramp where the charge was injected.  However, it oversubtracts signal elsewhere, resulting in long negative tails that would extend along the track past the limits of the image shown.  This oversubtraction occurs because our moving object model cannot account for signal deposited all at once: a moving object model that deposits charge at a designated time must necessarily deposit charge continuously throughout the ramp.  The $\chi^2$ of the fit, in the lower-right panel, is poor, with many pixels having $\chi^2$ values $\gtrsim$50 with four degrees of freedom.

Oblique cosmic rays may sometimes look similar to the left panel of Figure \ref{fig:romancal_comparison} if jump detection does not completely remove them.  Such a track could then be flagged for further processing, which would need to distinguish a true moving source from an oblique cosmic ray.  Figures \ref{fig:romancal_comparison} and \ref{fig:oblique_CR} show that our methodology can clearly distinguish between the track of a moving object, which can be well-fit, and an oblique cosmic ray, which cannot be.  

\section{Achievable precision and the impact of the read pattern} \label{sec:precision_read_pattern}

Section \ref{sec:fitramp} presents two approaches to fit the track of a moving object: one if the static astronomical scene is unknown, and a simpler, more precise approach if it is known.  In this section we show the precisions achievable in each case.  Legacy approaches to fitting tracks \citep{Veres+Jedicke+Denneau+etal_2012,Virtanen+Poikonen+Santti+etal_2016,Bouquillon+Mendez+Altmann+etal_2017} do so only in integrated light.  In the best possible case this would correspond to a known static scene, e.g., a constant background.

For the joint fit of a moving object and a static scene to work, a source should move by several resolution elements during an exposure.  An asteroid 1\,au from Earth moving at 15\,km\,s$^{-1}$ relative to Earth would move $\approx$0.$\!\!''02$\,s$^{-1}$, or $\approx$0.6\,pixels/read given the readout time of Roman-WFI \citep{Schlieder+Barclay+Barnes+etal_2024}.  The motion of such a near-Earth asteroid should be easily resolved by Roman.  A Kuiper belt object would have a similar relative velocity to Earth but might lie 40\,au or 50\,au away, resulting in motion of just $\sim$0.01\,pixels/read.  Over 30 reads this motion is a fraction of a resolution element: we do not expect to be able to effectively fit the sky paths of icy outer bodies unless the static scene is known.  

A similar resolution criterion applies to the inference of size and shape.  Assuming a source to lie at a distance of 1\,au from the Roman space telescope, a diameter of one Roman-WFI pixel corresponds to nearly 100\,km.  For the vast majority of asteroids, we therefore do not expect to be able to obtain direct measurements of their sizes and shapes.  Nonuniform albedos and complex shapes, common in small asteroids, further complicate any prospects of measuring moving objects' shapes.  The analysis in this section will therefore take the moving objects to be point sources.  

For a wide-field, ground-based observatory with the ability to read out up-the-ramp, a terrestrial satellite would move by upwards of $\approx$200 arcseconds/second, i.e., by many resolution elements in a single read.  The ground-based case provides an extreme example of rapid source motion that can be addressed with the techniques in this paper, assuming that a sufficiently rapid up-the-ramp readout scheme is available.

\subsection{Flux Precision: Known Static Scene}

We begin with the case of a known static scene, which we take to be a constant background.  This is the ideal scenario for fitting a moving object.  There is no requirement that the object move by several resolution elements during the exposure: the small-velocity limit corresponds to PSF photometry.  In our approach this corresponds to Equations \eqref{eq:b_knownbg} and \eqref{eq:err_b_knownbg}.  

To test the precision that is possible with a perfect source+scene model, we subtract the (known) expected background and apply Equation \eqref{eq:err_b_knownbg} to compute the uncertainty in the moving object's flux.  We consider three different readout patterns: 32 reads grouped into six resultants as described in Section \ref{sec:synthetic_data}; 32 reads saved individually; and two reads only, with only the total counts available.  The full approach with an unknown static scene cannot be applied in the latter case because there is no time resolution in the exposure.

\begin{figure*}
    \centering
    \includegraphics[width=0.45\linewidth]{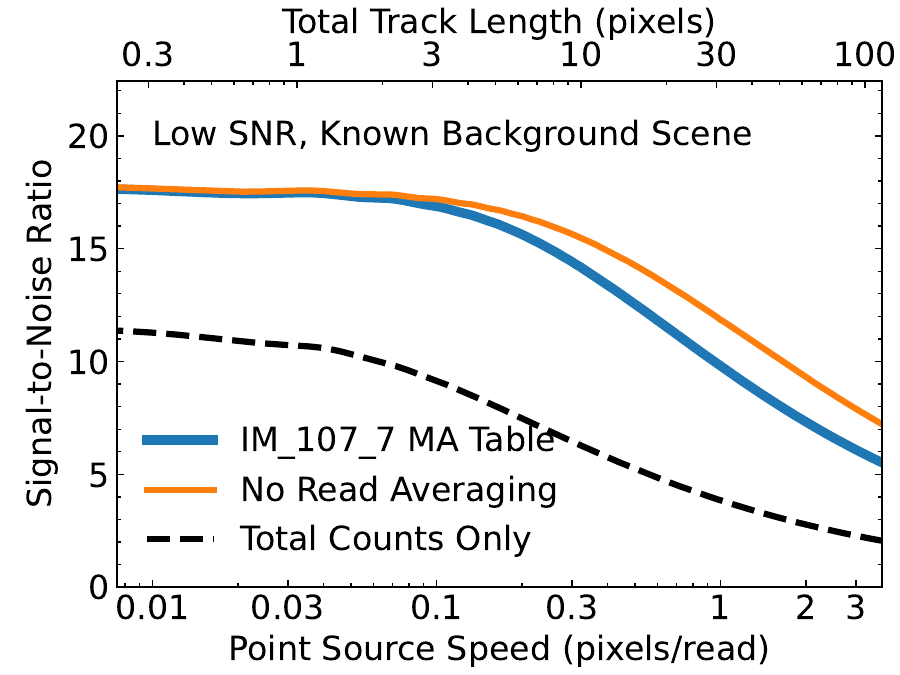} \hspace{0.1 truein}
    \includegraphics[width=0.45\linewidth]{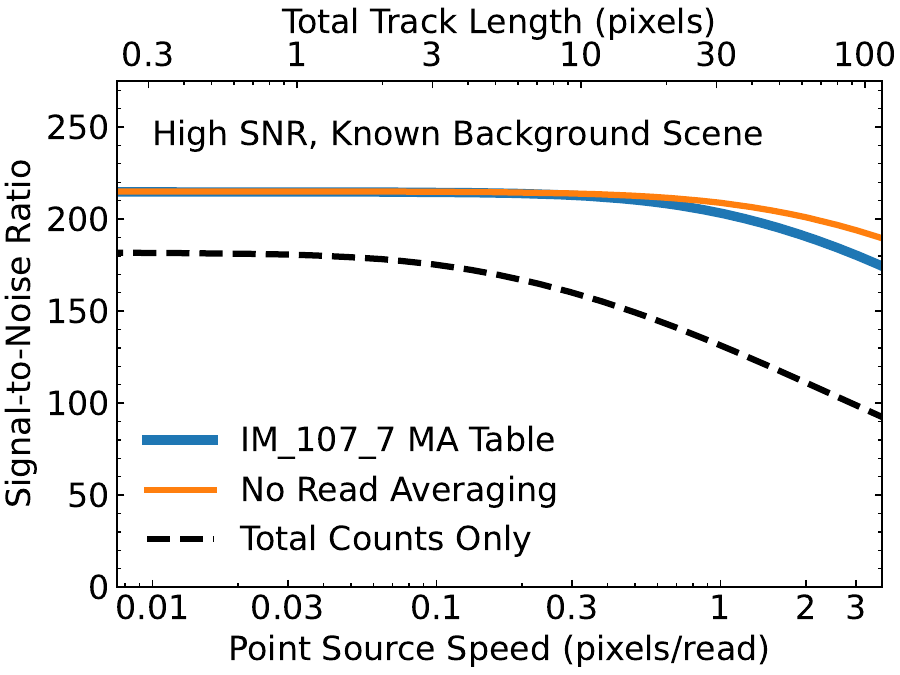}
    \caption{Signal-to-noise ratio of a moving point source as a function of its speed across the detector for a faint source (left) and a bright source (right).  The calculation assumes that both the path of the moving object and the static astronomical scene are known.  The thicker blue lines use the same readout pattern as Figure \ref{fig:modelresultants_hlwas}; the thinner orange lines assume the same 32 reads but with each read saved individually.  The black lines show the precision achievable from fitting a track only to the integrated counts, i.e., to the difference of the first and last reads.  At low speeds, the fit is equivalent to PSF photometry.}
    \label{fig:snr_speed_knownbg}
\end{figure*}

Figure \ref{fig:snr_speed_knownbg} shows the results.  The signal-to-noise ratio (SNR) of the fainter source, shown at left, is limited by the background; it would have an SNR of $\approx$33 in the absence of any background or read noise.  The bright source shown at right is photon noise limited.  The thicker blue curve adopts the resultant averaging discussed in Section \ref{sec:synthetic_data} while the thinner orange line assumes every read to be available individually.  

In the left image, read noise (taken to be 10\,e$^-$/read) is comparable to shot noise from the 5\,e$^-$/read background; the use of many reads offers significant sensitivity gains.  As speed increases, the background photons only add noise for the duration of the moving object's passage across that pixel so long as this can be resolved by the readout pattern.  As a result, the use of the full ramp offers large sensitivity gains over simply fitting the integrated counts.  The loss of sensitivity at high speeds arises because the readout pattern still results in smeared resultant-to-resultant differences (c.f.~Figure~\ref{fig:ePSF_smearing}), increasing the importance of background photon noise.  The readout pattern averaging reads into resultants is moderately less sensitive than the case of saving all of the reads.  

\subsection{Flux Precision: Unknown Static Scene}

In the general case we do not know the background astronomical scene.  In order to jointly fit the scene and a foreground moving object, that object must move appreciably over an exposure.  We therefore restrict our analysis to the precisions obtainable for unresolved moving sources with speeds ranging from 0.01 to 3\,pixels/read.  As in the previous subsection, we adopt a uniform background for simplicity, though we no longer impose this knowledge when fitting.  Equation \eqref{eq:err_b_knownbg} provides an uncertainty on the flux of the point source that will be accurate if the path and position are known.  This allows us to measure the ideal signal-to-noise ratio (SNR), the moving-object equivalent of forced photometry.  
\begin{figure*}
    \centering
    \includegraphics[width=0.45\linewidth]{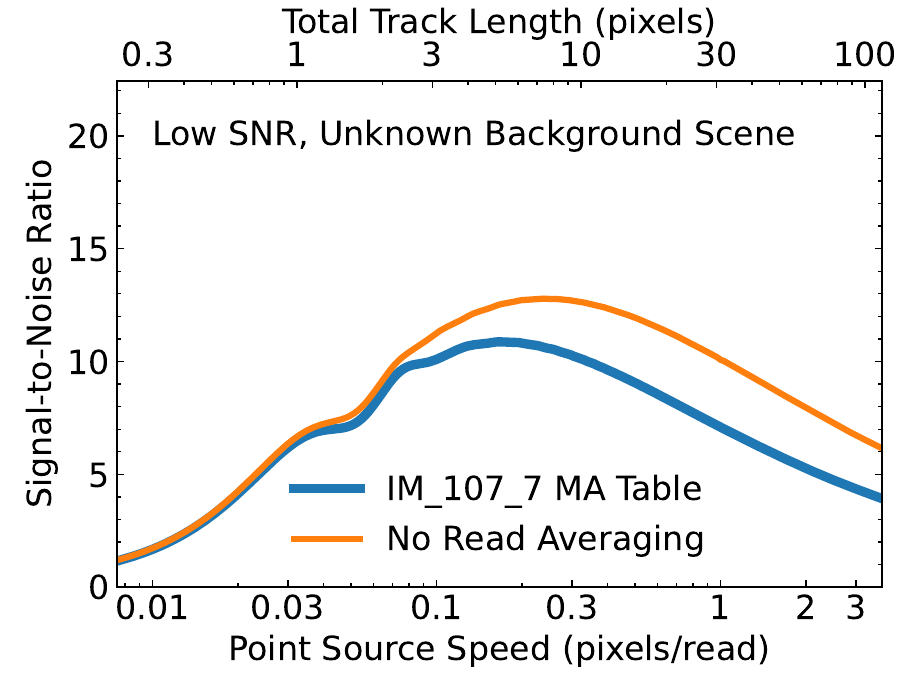} \hspace{0.1 truein}
    \includegraphics[width=0.45\linewidth]{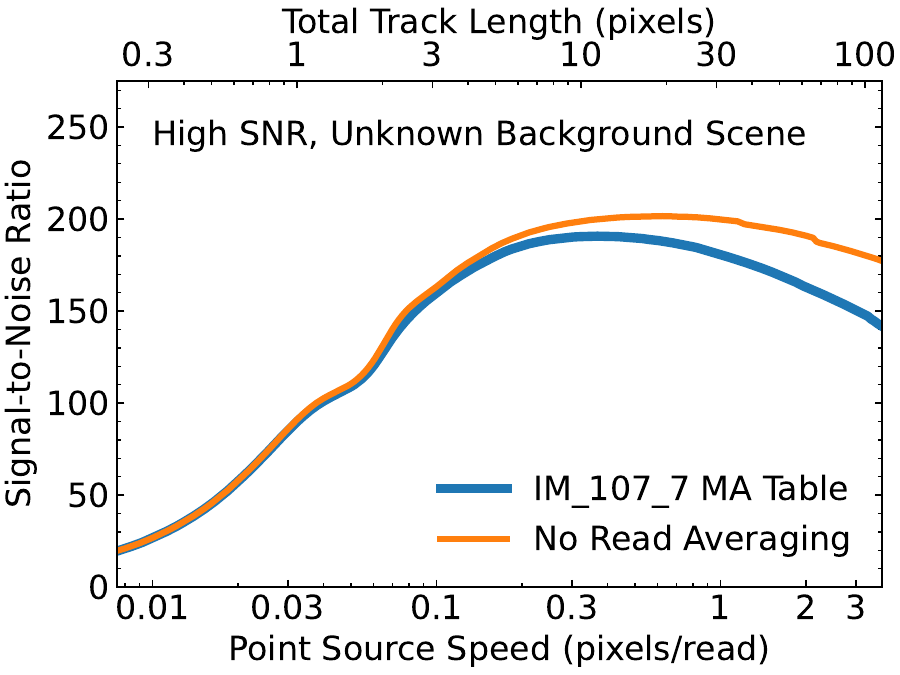}
    \caption{Similar to Figure \ref{fig:snr_speed_knownbg}, but  assuming that only the moving object's path is known; the static astronomical scene must be fit pixel-by-pixel.  At low speeds, it becomes impossible to distinguish flux from moving and static sources, while at high speeds, the image is sufficiently smeared out within resultants that read and photon noise from the background increase and SNR falls. }
    \label{fig:snr_speed}
\end{figure*}

Figure \ref{fig:snr_speed} shows the SNR as a function of speed across the detector for the case of an unknown background scene.  At low speeds across the detector, it becomes impossible for our approach to distinguish flux from a moving object from flux from a stationary object.  Regardless of the moving object's brightness, its SNR drops sharply as its speed approaches zero.  If we are limited by the photon noise of the background, SNR also falls at higher source speeds as a more spread out ePSF (c.f.~Figure \ref{fig:ePSF_smearing}) incurs more background noise in each resultant difference.  This happens at lower speeds for the fainter source.  

For real observations, the position and speed of a moving source are unlikely to be known well enough to apply the equivalent of forced photometry.  In this case we must also fit for the position and velocity.  At low SNR there is a risk of catastrophic failures in such a fit, and the prospects for detection will be worse than implied by Figure \ref{fig:ePSF_smearing}.  For bright, relatively fast-moving sources, this is much less of a concern.

\subsection{Precision of Position and Velocity}

It is more difficult to estimate the precision of a position or speed measurement than it is for a flux measurement.  We proceed empirically: we generate many noise realizations from a single sky path and measure the scatter of the best-fit position and velocity.  For simplicity we take the true sky path to be aligned with one of the detector axes.  We generate 10000 noise realizations for the sky path at cross-detector speeds ranging from 0.1 to 3 pixels/read.  In all cases the total flux from the source is fixed; the integrated signal-to-noise ratio of the track depends only weakly on the speed across the detector (c.f.~Figure \ref{fig:snr_speed}).  Throughout this section we adopt a source flux of 100~e$^-$/read, a background level of 5~e$^-$/read/pixel, read noise of 10~e$^-$, and the 32-read, six-resultant readout pattern discussed in Section \ref{sec:synthetic_data}. 

\begin{figure}
    \centering
    \includegraphics[width=\linewidth]{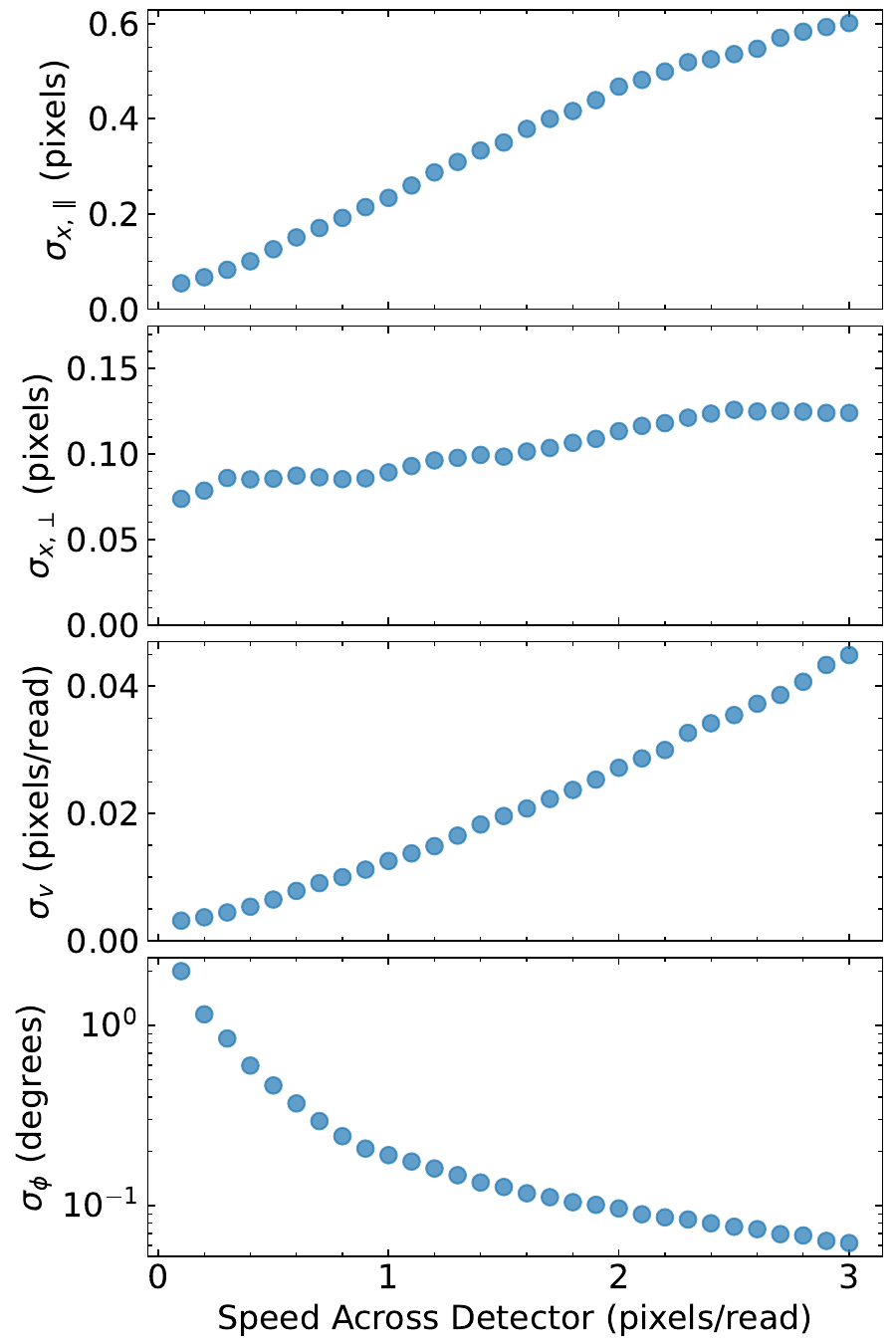}
    \caption{Precisions in position and velocity for a moving source with flux of 100~e$-$/read in a 32-read, six-resultant ramp, computed using Monte Carlo over noise realizations.  From top to bottom, we show the precision on position parallel and perpendicular to the track, the speed, and the direction of motion.  }
    \label{fig:precisions_position_velocity}
\end{figure}

Figure \ref{fig:precisions_position_velocity} shows our results for the four parameters describing position and velocity.  The precision of the transverse position is insensitive to the track length at a similar integrated SNR.  The reference position parallel to the track, however, is much less precise at faster track speeds.  The precision of the speed measurement itself also falls with increasing speed.  In this case it remains a relatively constant fractional value of $\approx$1.5\%.  The precision on the track orientation, and hence on the angle of the moving object's path across the detection, improves steeply with speed (note the logarithmic axes in the bottom panel).  Increasing the flux, and hence the SNR, of the source would improve all precisions shown.  

The uncertainties shown in Figure \ref{fig:precisions_position_velocity} are computed using Monte Carlo rather than from a quadratic form approximation to $\chi^2$.  The use of bicubic splines for the ePSF, the subsequent required integrations, edge effects from the finite size of the ePSF, and an iterative flux calculation to mitigate some forms of bias can all contribute small, discrete perturbations to $\chi^2$.  In our tests, using finite differencing of $\chi^2$ to estimate the covariance matrix did not produce reliable results.  If a user wishes to estimate their uncertainties, we recommend generating many noise realizations from the best-fit sky path, fitting them, and computing summary statistics on those fits.  Alternatively, approaches like Markov Chain Monte Carlo can directly sample from the posterior distribution of the parameters.  Both of these approaches to estimating uncertainties are much more computationally intensive than performing a maximum likelihood fit.

\section{Biases in the Fitted Parameters} \label{sec:biases}

Section \ref{subsec:maxlike} briefly discussed the mitigation of bias on the recovered flux due to covariance of estimated and realized photon noise.  In this section we assess biases of our approach more broadly.  We separate these into biases on the recovered flux and biases on the recovered astrometry---the position and velocity of a moving source.

We first estimate the bias by generating 10000 sample tracks oriented along one of the detector axes with a speed of 3 pixels/read and the read pattern used in Section \ref{sec:synthetic_data}.  As in the previous section, we adopt a source flux of 100~e$^-$/read, a background level of 5~e$^-$/read/pixel, read noise of 10~e$^-$, and the 32-read, six-resultant readout pattern from Section \ref{sec:synthetic_data}.  For each sample track, we use the Nelder-Mead algorithm to find the maximum-likelihood trajectory and flux of the moving object.  We find the recovered position and velocity of the injected source to be unbiased assuming an accurate ePSF.

It is more difficult to derive an unbiased flux than to derive an unbiased position and velocity.  \cite{Portillo+Speagle+Finkbeiner_2020} derived biases in flux when simultaneously fitting for a position and brightness of a source atop a flat background.  They found a small positive bias in the maximum likelihood flux, though this bias was absent in the flux when marginalized over the uncertain position.  

The calculation presented here is more complex than the scenario in \cite{Portillo+Speagle+Finkbeiner_2020}.  We are fitting for more parameters than there are pixels: a constant count rate for each pixel plus a count rate, a position, and a velocity for the moving source.  Unlike \cite{Portillo+Speagle+Finkbeiner_2020}, we make no assumption of a flat background.  We therefore expect their results to apply qualitatively but not quantitatively.  

\begin{figure}
    \centering
    \includegraphics[width=\linewidth]{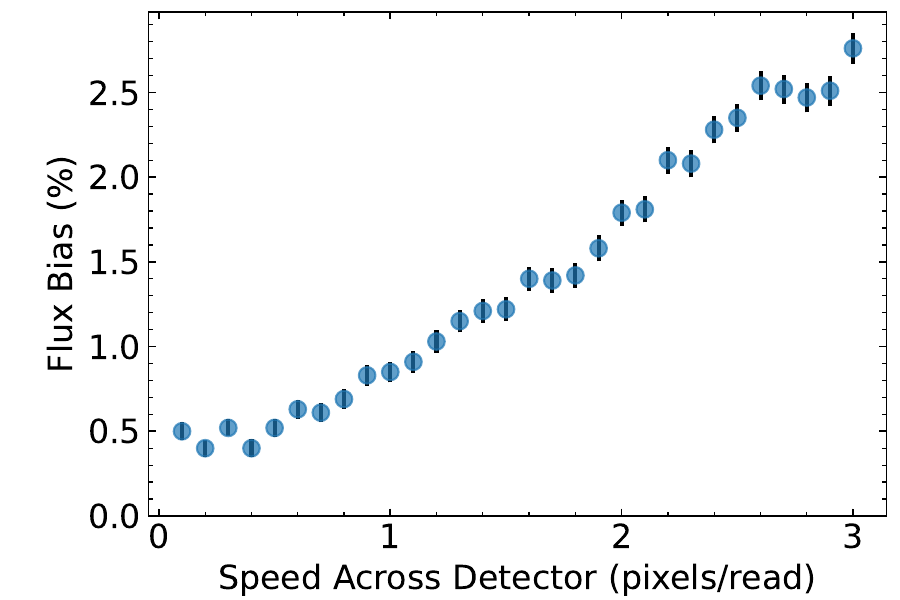}
    \caption{Bias in the maximum likelihood flux as a function of cross-detector speed for a source with a flux of 100 e$^-$/read in a 32-read, six-resultant ramp (total ${\rm SNR} \approx 20$).  A positive bias indicates that the recovered flux is, on average, larger than the input (true) flux. }
    \label{fig:fluxbias}
\end{figure}

Figure \ref{fig:fluxbias} adopts the same approach as Figure \ref{fig:precisions_position_velocity} to compute the precision and bias of the flux: we generate 10000 noise realizations for tracks at a range of speeds across the detector.  The error bars in Figure \ref{fig:fluxbias} represent the scatter in recovered flux values divided by $\sqrt{10000}$, i.e., the square root of the number of Monte Carlo realizations.  Throughout the calculation we adopt the readout pattern described in Section \ref{sec:synthetic_data} consisting of 32 reads grouped into six resultants. and we hold the true flux constant.  As the object's speed increases, its flux is spread out over a longer track on the detector.  As the speed increases, we find that our precision on flux degrades in agreement with Figure \ref{fig:snr_speed}, but we also find an increasing positive bias.  This is in line with the findings of \cite{Portillo+Speagle+Finkbeiner_2020}, but the dependence on speed reflects the increasing number of pixels---and therefore best-fit static count rates---that are relevant to the optimization.  At all source speeds, the bias in flux is smaller than its uncertainty in any given noise realization, typically by a factor of at least $\approx$3.  

We next assess whether the bias in Figure \ref{fig:fluxbias} may be avoided by marginalizing rather than optimizing over position and velocity.  For this step we take only a single speed of 3 pixels/read and marginalize over position and speed for each of 1000 noise realizations of the same underlying track.  We find that the bias of $\approx$3\% shown in Figure \ref{fig:fluxbias} falls, but to $\approx$1.3\% rather than to zero.  In other words, marginalizing over position and velocity will reduce the bias on the recovered flux, but this procedure may not eliminate it.  If the intrinsic track is accurately known, e.g.~from an independent orbital fit, then the bias on the recovered flux may be avoided entirely.  As for the uncertainties in the position and speed discussed in Section \ref{sec:precision_read_pattern}, a user wishing to perform precision photometry on a moving source may need to use Monte Carlo to estimate the biases at the $\sim$1\% level.

Biases in a moving object's flux can affect science in two main ways: by providing a slightly incorrect measure of brightness, and by providing an overly bright track that is subtracted from a noisy image before subsequent processing.  The latter effect could be significant if our methodology is applied to ground-based data with bright tracks from terrestrial satellites.  These satellite tracks likely contain no scientific value, but their removal may introduce small biases in the recovered value of the underlying sky scene.  

\section{Computational Implementation and Performance} \label{sec:implementation}

We have implemented the moving object fitting approach in this paper in Python; the code is available at \url{https://github.com/t-brandt/moving_source}.  There are three computationally demanding components: the calculation of the ePSF as smeared over a single read difference; bicubic spline interpolation of that ePSF for every read difference, and the calculation of the per-pixel likelihood.  The former two are written in Python using {\tt scipy.interpolate.RectBivariateSpline}, while the latter is written in Cython.  A Cython implementation is a factor of several faster than a Python implementation following that of \citetalias{Brandt_2024}.

The main way for a user to interact with the Python module is via the {\tt MovingTrack} class.  This class can be used to generate a moving object's track from a position, speed, readout pattern, and ePSF.  It can also fit a track using a series of resultants, a readout pattern, an initial guess for the position and speed, and an ePSF.  We do not include jump detection in the fit as it would mask the resultant differences with the highest SNR of the moving object.  If a user wishes to mask resultant differences this can be done by passing a boolean mask to the fitting method of {\tt MovingTrack}.  

Fitting a moving object's track requires minimizing $\chi^2$; this is done through {\tt scipy.optimize.minimize}.  The user may pass any minimization method that is accepted by that function.  For this paper we have found good performance with Nelder-Mead.  With a good initial guess and four nonlinear parameters, the optimization typically requires several hundred computations of a moving object track followed by evaluations of the per-pixel $\chi^2$.  This process takes a few ms per iteration, or $\sim$1 second in total on a 2023 Macbook Air assuming that the fit is done in a $60 \times 70$ pixel cutout.  Computational effort increases with cutout size, ePSF extent, and the number of both reads and resultants.

We implement two further approximations to improve computational performance.  First, we compute $\chi^2$ for every pixel without fitting a moving object, and replace this with the moving object $\chi^2$ only for those pixels where the moving object track contributes a minimum amount of signal (by default $10^{-4}$ of the peak signal).  This substantially reduces the number of pixels that must be treated using the approach of Section \ref{sec:fitramp} without significantly affecting the result.  Our second approximation is to use a first-order Taylor expansion for the smeared ePSF in the case where the optimization routine makes a very small change to the velocity.  By default we use this expansion when the increment is at most 0.05 pixels over a read.  This approximation becomes relevant when the optimization routine is approaching convergence and takes very small steps.  Our use of the two approximations listed here approximately halves the time to fit the track of a moving object.  

\section{A Demonstration on JWST Data} \label{sec:jwst_data}

We now test the approach described in this paper on data from the JWST NIRISS instrument \citep{Doyon+Willott+Hutchings+etal_2023}.  We use a cutout around a moderately bright asteroid identified by \cite{Martel_2025} in an F115W image from Program 1571, PI Malkan\footnote{Observation ID {\tt jw01571102001\_06201\_00001}}.  The data consist of a single integration with 12 groups of four reads each.  We process the data using the JWST pipeline \citep{JWST_pipeline_2023} up to the ramp fitting step, using the argument {\tt --save\_calibrated\_ramp} to preserve the groups before ramp fitting.  We do not use the pipeline's jump flags for most of our fits, though we do use them to compute $\chi^2$ values for the pipeline fit.  Our method requires the gain (to compute photon noise), the flatfield (to multiply the intensity of a moving source), and the read noise; we use the appropriate files available on CRDS\footnote{\url{https://jwst-crds.stsci.edu}}.  We convert all units to electrons by multiplying by the gain, we apply the same conversion to the read noise, and we convert back after fitting a moving source to enable a direct comparison to the JWST pipeline outputs.  We use the ePSF for this instrument and band as derived by \cite{Libralato+Bellini+vanderMarel+etal_2023}.  

\begin{figure}
    \includegraphics[width=\linewidth]{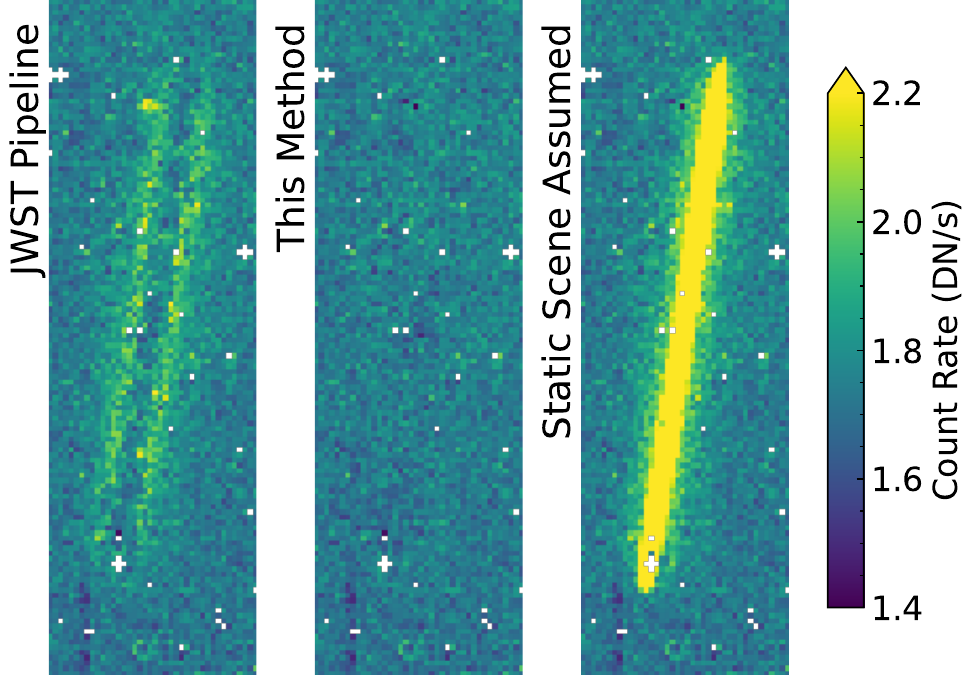} \\[1em]
    \includegraphics[width=\linewidth]{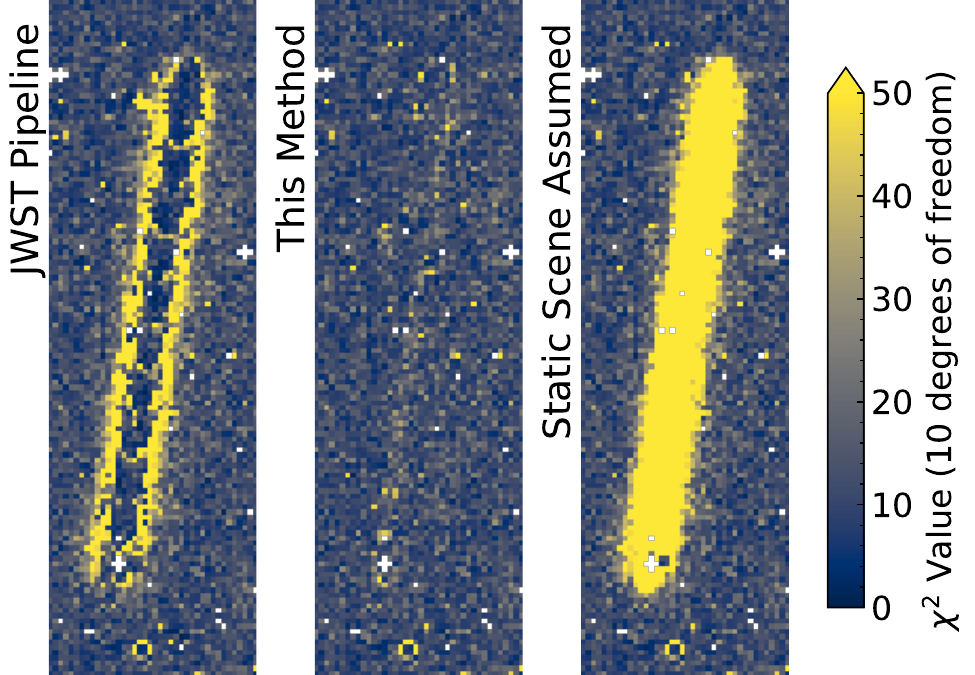}
    \caption{Demonstration of the algorithm in this paper on an asteroid track observed by JWST/NIRISS.  The top-left panel shows the {\tt rate} image produced by the JWST pipeline, the top-middle image shows the static count rate we infer with our approach, and the top-right panel shows the count rate we would infer assuming a static scene and without rejecting the asteroid track as jumps.  The lower panels show the per-pixel $\chi^2$ values for each fit shown at top. }
    \label{fig:jwst_data}
\end{figure}

The top panels of Figure \ref{fig:jwst_data} show the inferred count rates as produced by the JWST pipeline as well as by our approach; the bottom panels show the $\chi^2$ values of these fits.  The top-left panel of Figure \ref{fig:jwst_data} shows the {\tt rate} image as produced by the JWST pipeline \citep{JWST_pipeline_2023} and downloaded from MAST\footnote{\url{https://mast.stsci.edu/search/ui/\#/jwst}}.  The core of the asteroid track has been flagged as jumps in the count rate and its signal has been been effectively masked from the ramp fit.  The outer parts of the track remain, as those jumps were not large enough to surpass the jump detection threshold.  The lower-left panel of Figure \ref{fig:jwst_data} shows the per-pixel $\chi^2$ value of the fit when omitting group differences flagged as jumps by the JWST pipeline.  The pixels at the edges of the track, which have visible artifacts in the top-left panel, have $\chi^2$ values $\gtrsim$50 for 10 degrees of freedom (11 group differences with one count rate per pixel).  Such a $\chi^2$ value is equivalent to a $5\sigma$ outlier under Gaussian statistics.  

The top-middle panel of Figure \ref{fig:jwst_data} shows the static count rate we infer from fitting a moving object track.  To remove cosmic rays, we have fit the asteroid's track twice.  With the first fit we obtain a model of the signal contributed by the asteroid for each group.  We then perform a simple jump detection on the groups after subtracting the modeled signal of the moving object: we mask group differences that differ from the median in a given pixel by more than 5 times the expected noise.  This noise represents both read noise and photon noise; the expected photon noise is derived from the sum of the static scene and the modeled moving object track in a given group difference.  We multiply the photon noise by $\sqrt{2}$ to avoid masking the core of the track where group averaging changes the calculation of the variance (\citetalias{Brandt_2024}).  We verify that our approach does not flag jumps along the track center.  We also mask bad pixels that are flagged as {\tt NaN} in the JWST pipeline data products.  After flagging jumps, we refit the moving object's track to obtain the best-fit static scene shown in the top-middle panel.  The lower-middle panel shows the resulting $\chi^2$ values.  There are a handful of pixels along the track with slightly elevated $\chi^2$ values.  These could indicate an imperfect (albeit still very good) ePSF, and/or slight photometric variability of the asteroid. 

The top right panel of Figure \ref{fig:jwst_data} shows the same scene as the other panels, but fitted using only the jumps that we flagged as described above and assuming only a static scene.  In this case the full track of the asteroid is clearly visible and manifests as a streak in the image.  This panel closely resembles what the JWST pipeline would produce if it could distinguish jumps resulting from a moving object from jumps due to cosmic rays and mask only the latter.  The lower-right panel shows the corresponding $\chi^2$ value from fitting the ramps assuming a static scene.  Without masking group differences representing the core of the moving object's track, the $\chi^2$ values can approach $10^5$ for 10 degrees of freedom.

The top middle panel of Figure \ref{fig:jwst_data} shows the inferred underlying scene when fitting a moving object.  In applying the methodology developed in this paper, we also obtain parameters for the asteroid itself: we derive an angle of $188.\!\!^\circ34$ degrees clockwise of vertical and a speed of 2.173 pixels/group (0.05059 pixels/second).  These values change by a few parts in $10^4$ depending on whether or not we mask jumps.  The readout of the detector also impacts the interpretation of this speed and, to a lesser extent, the direction of motion.  The difference in the time of readout between the top and bottom of the track, which are separated by $\approx$100 pixels, is $\approx$0.5 seconds, or $\approx$10$^{-3}$ of the 515-second total exposure time.  This fractional effect is comparable to the precision attainable with our measurement, and would slightly modify our derived speed in the slow (vertical) readout direction.  We obtain a flux of $1650.6 \pm 1.5$ DN/second without masking jumps, or $1652.6 \pm 1.5$ DN/second with masking.  Masking of cosmic ray jumps allows for a slightly more aggressive fit of the moving object template to the data and better residuals.  As with the speed of the object, our derived flux will be modified at the $10^{-3}$ level (i.e.~comparable to our uncertainties) by the rolling readout of the detector.  

\section{Conclusions} \label{sec:conclusions}

This paper has presented a method for fitting a moving object in an astronomical image read out up-the-ramp.  The method maximizes the likelihood pixel-by-pixel to disentangle flux from a moving object from flux from an arbitrary but static astronomical scene.  The likelihood may be evaluated efficiently, enabling a fit of the moving object's path across the detector through the various reads.  This provides the best measurements of a moving object's position and velocity and enables its track to be removed from any static scene.

Our approach may be contrasted with fitting the track of a moving object only in integrated counts.  In that case, a track may only be cleanly fit when assuming the spatial structure of the background or when independent knowledge of that background is available, e.g., from difference imaging.  Our use of the individual reads also improves the precision with which position, velocity, and flux may be measured.  

Our maximum likelihood approach is unbiased in the recovered position and velocity, but it does have a small bias in the recovered flux.  This bias arises because the position and velocity may be optimized by perturbing the fit in the direction of a noise fluctuation.  The bias may be mitigated, but not fully avoided, by marginalizing over the uncertain position and velocity rather than by performing a maximum likelihood fit. 

Our approach requires a good initial guess for the position and velocity of a moving object in order for the optimization to converge.  Given a method for curating data, identifying likely tracks, and approximating their positions and speeds, our likelihood-based approach enables precise measurements independently of the background scene.  It may be applied to data from any space-based observatory reading out up-the-ramp, including HST, JWST, and Roman.  

We have demonstrated a successful application of our approach to JWST data from the NIRISS instrument.  With a good ePSF for the instrument, we obtain an excellent fit to an asteroid track.  At the level of precision attainable for this observation the rolling readout of the detector impacts our analysis at a level comparable to our measurement error.  

Our approach could also be applied to ground-based data; the moving objects in this case would mostly be terrestrial satellites.  These satellites, if in low-Earth orbit, could move at speeds of hundreds of detector pixels per second.  If the applicable detectors can be read out nondestructively at a sufficient cadence, the tracks of such satellites can be fitted and removed much better than in integrated light.  The proliferation of satellites in relatively low-altitude orbits could provide a strong incentive for the development and installation of such detectors.

\noindent {\it Software}: 
          scipy \citep{2020SciPy-NMeth},
          numpy \citep{numpy1, numpy2},
          matplotlib \citep{matplotlib},
          astropy \citep{Astropy_2013,Astropy_2018,Astropy_2022},
          Cython \citep{Cython},
          Jupyter (\url{https://jupyter.org/}).

%\begin{acknowledgements}
%The acknowledgements.
%\end{acknowledgements}

\bibliography{movingsource}{}
\bibliographystyle{aasjournal}

\end{document}